\documentclass[manuscript]{aastex}

\usepackage{color}

\slugcomment{}

\shortauthors {Stanghellini \& Haywood}
\shorttitle {Planetary nebulae and Galactic evolution} 

\begin{document}

\title {Galactic planetary nebulae as probes of radial metallicity gradients and other abundance patterns.}

\author{Letizia Stanghellini}
\affil{National Optical Astronomy Observatory, Tucson, AZ 85719}
\email{lstanghellini@noao.edu}

\and
\author{Misha Haywood}
\affil{GEPI, Observatoire de Paris, CNRS, Universit\'e Paris Diderot, 92190 Meudon, France}
\email{misha.haywood@obspm.fr}

\begin{abstract} 
We use planetary nebulae (PNe) as probes to determine the Galactic radial oxygen gradients, and other abundance patterns. We select data homogeneously from recent data sets, including PNe at large Galactocentric distances. The radial oxygen gradient calculated for the general PN population, which probes the region between the Galactic center and out to $\sim$28 kpc, is shallow, with slope $\sim$-0.02 dex kpc$^{-1}$, in agreement with previous findings. 
We looked for time evolution of the metallicity gradient using PNe with different age progenitors as metallicity probes. We identify PNe whose progenitor stars are younger than 1 Gyr (YPPNe), and those whose progenitor stars are older than 7.5 Gyr (OPPNe), based on the comparison between evolutionary yields and elemental abundances of the PNe.  
By studying OPPNe and YPPNe separately we found that: (i) The  OPPNe oxygen gradient is shallower ($\sim-0.015$ dex kpc$^{-1}$) than that derived from YPPNe ($\sim-0.027$ dex kpc$^{-1}$); (ii)  the OPPNe inner radial distribution of oxygen is compatible with no gradient to the radial extent of the thick disk population ($\sim$10 kpc), similarly to what has been observed in thick disk stars; (iii) planetary nebulae (especially OPPNe) indicate that significant gradient slope is limited to Galactocentric distances between $\sim$10 to $\sim$13.5 kpc, as observed for open clusters and field stars. Outside this range, the distribution is almost flat.
We found that the radial oxygen gradient is steeper for a PN population closer to the Galactic disk, similarly to what is observed in the general stellar population by the SEGUE survey. 
We use our novel population dating to compare our results with current chemical evolutionary models, and with gradients from other Galactic populations, for insight on galaxy chemical evolution. 

\end{abstract}


\section {Introduction} 
Radial metallicity gradients have been successfully used in the recent past, and they are still used, to set important constraints on the chemical evolution of galaxies. Some fundamental key questions, however, are still under debate. A wealth of recent generation data sets from Galactic and extragalactic surveys have brought new information on several aspects of metallicity gradients in star-forming galaxies, including how their shape varies with galaxy properties, whether they can be {multi-slope, and their evolution with time. If we limit gradient analysis to oxygen gradients, we still have a large variety of behaviors, and we seem to be far from a clear consensus on this essential constraint to chemical evolutionary models of galaxies.

Planetary nebulae are asymptotic giant branch (AGB) stellar ejecta; in turn, AGB stars are evolving from main sequence stars whose masses are in the 1-8 M$_{\odot}$ range. It has been often assumed that oxygen (and other $\alpha$-element such as neon and argon) abundances are invariant in low- and intermediate-mass stellar evolution. \cite{ventura2017} have compared AGB yields to Galactic PN abundances, using the most recent AGB evolutionary models, and an homogeneous and updated set of high quality abundances. While the initial abundances of low- and intermediate-mass stars in models that are built to reproduce the Milky Way stars are typically assumed to be solar, it is clear that $\alpha$-element abundances span a considerable domain. The comparison by \cite{ventura2017} shows that most observed Galactic PNe have progenitors with metallicity in the 0.008$<$Z$<$0.04 range (Z$_{\odot}\simeq0.018$ Asplund et al. 2009) . The final yields of these stars depends on the progenitor masses as well as metallicity. Carbon stars eject their shells and produce carbon-rich PNe, while stars that undergo the hot-bottom burning (HBB), where most of their carbon is transformed into nitrogen, do not go through the carbon star phase, and eject a nitrogen-rich shell. In Figure~1 (right panel) of \cite{ventura2017} we see that oxygen abundance varies with evolution noticeably only for Z$<$0.008, with a possible exception of HBB processing stars with Z$\sim$0.008. We can thus assume that oxygen is roughly invariant for AGB evolution in the Galaxy. On these premises, it is possible to use oxygen abundances of PNe to trace the oxygen abundance of the progenitor's population, and that radial (or vertical) oxygen gradients from PNe trace the gradient at the time of PN progenitor's formation, modulo subsequent dynamical evolution. In turn, the time of formation of the PN progenitor depends on the mass of the progenitor itself. By sorting PNe into progenitor age groups, and derive metallicity gradients for different age groups, we could determine the evolution of oxygen gradient in the Galaxy. 

Higher oxygen abundance in PNe located toward the Galactic Center has been noted since the 70s \citep{dodorico76}, based on a handful of objects. Until recently,  it has been difficult to discern whether the gradient was due to actual oxygen variations within the Galactic disk, or to detecting enhanced metals in specific populations such as the bulge, as shown by \cite{kaler1980}. \cite{maciel2003} compared the radial oxygen gradients from PNe to simple chemical evolutionary models of the Galaxy.  By using exclusively direct abundances (i.e., those obtained by determining electron density and temperature of each target via plasma diagnostics) of Galactic PNe, most authors have measured shallow gradients, with oxygen abundances higher near the Galactic center. 

\cite{SH10} (hereafter SH10) have determined that the general Galactic PN gradient is shallow (-0.023 dex kpc$^{-1}$) and may steepen with Galaxy evolution, demonstrating this with samples of PNe with different progenitor ages. While PN progenitor dating and distances carry uncertainties, to date there is still not a data set that contradicts this result. In this paper, we propose a revised analysis of the SH10 results, since several data sets of PN abundances, especially at the Galaxy anti-center, have become available in the last decade. Here, we also propose a novel method of PN progenitor dating, and an updates comparison with different data sets.

In $\S$2, 3, and 4 we describe the choices of PN distances, abundances, and progenitor ages. Gradients for the general sample are described in $\S$5, and time evolution of the radial gradients are discussed in $\S$6. We also consider vertical variation of $\alpha$-elements  gradients ($\S$7), and PN morphology ($\S$8). Finally, we discuss our results, in $\S$9, and draw the conclusions in $\S$10. 

\section{Distances and radii of the PNe}

To select the database for the present analysis, in \cite{Acker92}'s PN catalog, we select all Galactic PNe with measured angular radii. To these, we add a new set of $\sim$50 compact (apparent diameter $<4 \arcsec$) Galactic PNe whose WFC3/{\it HST} images have recently become available \citep{stanghellini2016}. 
The Galactic PN distance scale scenario is pretty much unchanged since 2010. \cite{stanghellini2008}'s (hereafter SSV) distance scale is still widely used in the literature, while a recent re-calibration of the surface brightness vs. physical radius scale has been published \citep{frew16}.  We use mostly the SSV scale in this paper, but also calculate some of the key gradients with other distance scale, aware that the SSV distance scale may produce long (heliocentric) distances for the largest PNe. 

\cite{stanghellini2017} have studied trigonometric parallaxes of central stars (CSs) of Galactic PNe within the initial Gaia release (DR1), where the comparison was too limited to assess whether the SSV or other distance scales were to be preferred.  A preliminary study of the DR2 PN parallaxes (Stanghellini et al. 2018, in prep) suggests that reliable parallaxes ($\sigma_{\rm p}<p$ and $p>0$) correspond to Galactocentric distances  between 5 and 10 kpc from the Galactic center, thus they are inadequate for the study presented in this paper.

There are now $\sim$900 Galactic PNe whose apparent diameters, 5 GHz (or H$\beta$) fluxes, and thus statistical distances are available, including the recent compact PN sample. We use all these PNe in the general sample, excluding those whose angular diameter is an upper limit in Acker's catalog.

\section{Abundances}

The major goal of this paper is to provide radial metallicity gradients based on a large sample of PNe, including data published  since our previous analysis in 2010. 
The most important physical parameters in our study are the elemental abundances of the PNe, calculated via direct method. In SH10, we used abundances from selected references, following the selection criteria described by \cite{perinotto2004}. Here, as in SH10, we also select abundances from \cite{KB1994}, \cite{perinotto2004}, \cite{stanghellini2006}, \cite{PBS2006}, \cite {KH2001}, \cite{milingo2002}, \cite{K2003}, and \cite{costa2004}. 

We made the sample homogeneous by using the same Ionization Correction Factor for all abundances, as in SH10, and we reduce all other methods to that of \cite{KB1994}, which is the most commonly used. it is worth noting that the different ionic abundance sets used plasma diagnostics based on different atomic data sets. While recalculation of all ionic abundances with a uniform atomic data set is beyond the scope of this paper, we have checked the impact of possible differences across the sets by recalculating a few ionic abundances by \cite{KB1994}, but using te newer ionic data set in {\it nebular} within {\it IRAF}. We found typical [O III] abundance differences of the order of 0.01 dex, with some larger divergences of 0.05 dex. This made us confident in our results, but of course there may be outliers if we were to recalculate all abundances.

To update the initial sample, we also included data sets published after 2010 that follow the same quality criteria. We thus added abundances from \cite{henry10}, who privileged as targets those PNe in the anti-center direction, building a small but essential new abundance sample. We included the work by \cite{dufour15}, who used space-based spectra to determine excellent abundances of a small sample of extended PNe. Finally, we also include abundances by \cite{GHG14}, who published a newly observed sample of Galactic PN abundances, and also analyzed homogeneously several spectra published in the literature. 

We use abundances in two ways. First, we select the best abundance available in the literature for a given element and nebula, chosen among the pre-screened references listed above; the best dataset for each PN is typically the last published, but we did chose an older reference in cases where the abundance was deemed uncertain by the Authors.  We call this method S, for selected abundances. Note that all gradients in SH10 have been derived with this procedure. Second, since there are now enough high-quality unrelated abundance sets, we also use averages of published abundances (A method).

We use abundance uncertainties from the original references, and calculate the abundance dispersions to be used as uncertainties when we use A method. Individual uncertainties are given in different terms by the different Authors, and it is worth describing here how we did homogenized the different sets. Several Authors associate formal errors analysis to their abundances, and their error analysis includes temperature, density, abundances, and measured emission line uncertainties. From this type of analysis, \cite{stanghellini2006} calculate $\sim10\%$ uncertainties for O/H, N/H, and He/H, and 
$\sim20\%$ uncertainty for Ar/H and Ne/H.  \cite{perinotto2004} also give explicit uncertainties. \cite{dufour15} give accurate uncertainties to all measured elemental abundances. \cite{henry10} give accurate errors for the new measured abundances, which we used here (but they do not offer new error analysis for the other data that this group has published earlier, see below). Finally, \cite{GHG14} gives logarithmic uncertainties for the abundances, which we used directly in our plots and gradients.
On the other hand, \cite{K2003}, \cite{KH2001}, and \cite{milingo2002} do not give accurate abundances, but a general upper limit to the O/H uncertainty, $\sim3\times10^{-4}$, for all PNe in their set.  If we assume that this uncertainty is representative of the highest oxygen abundances measured in their work, we can conservatively assume uncertainties of 0.1 dex. Finally, \cite{KB1994}, \cite{PBS2006}, and \cite{costa2004}  do not explicitly mention abundance uncertainties. In these cases, we conservatively assumed uncertainties of 0.2 dex on the logarithmic  log(X/H)+12 abundances. 

\section{Ages of PN progenitors}

AGB stars, whose turnoff mass spans the $\sim$0.8 to 8 M$_{\odot}$ range, are the progenitors of PNe. The  upper and lower mass limits of this range are rather uncertain, the lower limit being set by observing that the dynamical vs. stellar evolutionary time must be such that the PN shell is illuminated by a hot star to produce a visible PN \citep{SR2000}, and the upper limit by the maximum turnoff mass of a CO core star. These critical masses vary with metallicity, as shown in several evolutionary model series \citep{ventura2016,ventura2017}. Since the ages of the progenitors in this mass range also cover a broad range, it is essential to find good ways to date the PN progenitors if we want to pinpoint the metallicity gradient at a certain time, or time interval, of the Milky Way evolution model, to compare these data with the models of chemical evolution. Dating PN progenitors is also essential if we want to compare our oxygen and other gradients with those in the literature, based on targets of different nature. 

The process of dating PN progenitors is complex. To this day, the best method to sort Galactic PNe in populations is by \cite{PTP83}: Type~I PNe are the progeny of the more massive AGB stars, Type~II are the commonly found intermediate mass disk PNe, and Type~III are the high velocity low-mass progenitor PNe. This classification is based on N/O (and He/H) abundance ratios, and on peculiar velocities of the PNe. 
The N/O ratio is the major diagnostics in PN Type selection. \cite{perinotto2004} shows that Galactic Type I PNe, i. e., those whose progenitors underwent HBB processing, have ${\rm log}(N/O)>-0.3$. This critical ratio changes with population metallicity; for instance, in Magellanic Cloud PNe the critical ratio for Type I PNe is ${\rm log}>-0.4$ \citep{Dop91}. 
The PN types method works reasonably well to identify young progenitors based on their chemistry and location in the Galactic plane, but not so well for the larger group of PNe with old progenitors ($t_{\star}>$7-8 Gyr, where $t_{\star}$ is the age of the stellar progenitor).

We can now take advantage of the many stellar abundances that have been published in the recent years, where the stellar spectra have been analyzed for both iron and $\alpha$-element abundances, and where ages and turnoff masses have been obtained by sequence fitting of the stars. AGB models have also advanced, and we use this wealth of new data and models to refine a novel way of dating PN progenitors. In $\S$4.1 we review the mass ranges of the progenitors of PN types, in $\S$4.3 we use the new stellar data to gain insight on PN dating. Finally, in $\S$4.3, we present our novel PN progenitor dating scheme, based on stellar evolution results.

\subsection{Mass and age ranges of the progenitors of PN types}

As described in SH10, Type~I PNe are those with log(N/H)+12$>$8.4 and log(He/H)+12$>$11.1; Type~II and Type~III PNe are defined as {\it non-Type~I PNe} with (absolute) peculiar velocities respectively smaller (Type~II) or larger (Type~III) than 60 km s$^{-1}$. Type~I PNe are probably evolving from AGB stars with mass $M>2-3 M_{\odot}$ \citep{ps80,Kaler1990}. Stars of these masses represent a young stellar population at all metallicities. Type~II are disk PNe, the progeny of the $1.2\le  M \le2 M_{\odot}$ stars at solar metallicity. Type~III PNe are those evolving from stars of $M\le1.2 M_{\odot}$ \citep{perinotto2004}. All these mass limits are very approximate, and they are valid only for Galactic PNe.

In Figure~1 we show the stellar-age vs. turnoff mass relation from \cite{maraston1998}'s fuel consumption theorem. The progenitor age ranges are $0 \le t_{\star} < 1$ Gyr for Type~I PNe, $1 \le t_{\star} < 5$ Gyr for Type~II PNe, and $t \ge 5$ Gyr for Type~III PNe, from their (very approximate) turnoff mass. In the figure, we mark the separations corresponding to the assumed limiting ages of the progenitors of Type~I, II, and III PNe. The stellar progenitors of Type~III PNe are old, but a sizable fraction of Type~II PNe may be old as well, thus it is difficult to sort Type~II and Type~III PNe into clear age groups.  The state of the art is that these PN Types do not offer a quantitative indication of their progenitor ages, and this is the reason why, in this paper, we are developing a more quantitative way to date PN progenitors.

\subsection{Insight from the old stellar population}

The extensive stellar data that has been flourishing in the recent years can give us more insight on the PN progenitor dating problem. \cite{ramirez2013} have analyzed hundreds of nearby FGK field stars, and derived their ages and masses, and several atomic abundances, including oxygen and iron.  We selected a subsample of their large stellar data set ($>$800 stars) to build an O/H vs. $t_{\star}$ relation that we can use to date PN progenitors. We started by studying their [O/Fe] vs. [Fe/H] relation, then scale it with their [O/Fe] vs. $t_{\star}$ relation.  We have only used stars with $5600  <T_{\rm eff} < 5900$ K, to keep only stars with the best determined 
atmospheric parameters .  We also selected stars with [O/Fe]$>$(-0.48[Fe/H]+0.04), in order to discard metal-poor thin disk stars, which are mainly objects of the outer disk seen in the solar vicinity, but which have a chemical 
evolution different from stars in the population that we want to compare to PN progenitors. 

The resulting sample is populated by 181 stars. We plot their ages and masses given in \cite{ramirez2013} in Figure~1. From the comparison of these data with the age-mass relation for old stars it is clear that Type~I and Type~II PN classes do not discriminate too well the progenitors ages.

In Figure~2 we plot log(O/H)+12, derived by us as described above, vs. stellar age. Clearly, stars with log(O/H)+12$>$8.7 have an almost equal chance to be younger or older than 7.5 Gyr. On the other hand, stars with log(O/H)+12$<$8.7 are almost invariably older than 7.5 Gyr.  Conversely, if we assume that oxygen is not produced nor processed by PN progenitors, and if we select PNe whose 
log(O/H)+12$<$8.7, we are selecting a very old disk population of PNe with progenitors older than $\sim7.5$ Gyr.  
This selection method works well to exclude young progenitors from PN samples whose oxygen abundance is known. By working with oxygen gradients we want to characterize the complete population of PNe with old progenitors, i.e., we like to include those PNe whose oxygen abundance is larger than 8.7, and whose progenitors have age$>7.5$ Gyr (i.e., the ones that represent the progeny of the stars in the upper-right corner of Fig.~2).  In order to do so, we need to look into AGB evolutionary yields, as described in the next section.

\subsection{Our PN progenitor dating scheme, based on AGB evolutionary yields}

During the AGB, the chemistry of the outer layers of the star change due to nucleosynthesis. These outer layers, once ejected and ionized, become the PN, thus abundance analysis of PNe can be directly compared to the final yields of AGB evolution, and from this comparison one can derive a sound assessment of the progenitor's mass and metallicity. The key elements for comparison between theoretical yields and observed abundances are carbon and nitrogen, largely processed in AGB stars, while oxygen and other $\alpha$-elements are almost invariant in this phase of stellar evolution, and are needed to frame the initial conditions of the models. Nitrogen is highly enhanced at the expenses of carbon in high-mass progenitors, thus PNe with enriched nitrogen have progenitor mass at least as high as the minimum critical mass for HBB processing.

While it is impossible to model the evolution of the progenitor of every single observed PN, in the last few years new models of AGB evolution have become available spanning a larger mass and metallicity domain.  \cite{ventura2016} show the final C/H, N/H, and O/H yields for a broad range of metallicities and for turnoff masses 
$1.5<M_{\rm TO}<8 M_{\odot}$.  We use their  N/H vs. C/H, and N/H vs. O/H, theoretical planes (Ventura et al. 2016, their Fig. 2) to determine zones where PNe are bound to have either old or young progenitors.

Old and young PN progenitors on the N/H vs. C/H plane occupy markedly different loci, thus carbon abundances, in combination with nitrogen abundances are the most useful to date PN progenitors. From Ventura et al. (2016, Fig.~2, right panel) we infer that if C/H$>$N/H the AGB star has gone through the carbon star phase. Thus, if we observe a PN with C/H$>$N/H we can be reasonably sure that it has a low-mass progenitor. On the other hand, if log(N/H)+12 is above the 0.57~log(C/H)+12+3.67 line (obtained by linear fit on the plot), the AGB star has gone through the HBB, thus is has $M>3M_{\odot}$, and $t_{\star}<$0.5-1 Gyr depending on metallicity.

Direct carbon abundances in PNe are based on the observation of bright carbon emission line in the UV, thus they are difficult to observe and need space-based data. Nitrogen and oxygen abundances are measured from optical  spectra.  To enlarge the sample of PNe for which we can determine the progenitor age we use the N/H vs. O/H diagnostics also. Following the plot by Ventura et al. (2016, their Fig.~2, left panel), we determine that PNe with nitrogen below the
log(N/H)+12=0.8~log(O/H)+12+1.4 line (fitted on the plot) are those that may descend from carbon stars. 
On the other hand, nitrogen abundances above the [log(N/H)+12]=0.6$\times$[log(O/H)+12]+3.3 line are the clear signature of the HBB occurrence, which indicates masses above $\sim$3 M$_{\odot}$,  thus  $t_{\star}<1$ Gyr. 

We define {\it PNe with old progenitors} (or OPPNe) those PNe with C/H$>$N/H, or,
[log(N/H)+12]$<$0.8$\times$[log(O/H)+12]+1.4.  We define {\it PNe with young progenitors} (or YPPNe) those PNe  whose abundances comply with the equations C/H$<$N/H, or [log(N/H)+12]$>$0.6$\times$[log(O/H)+12]+3.3. 

\subsection{The final sample}
Tables~1 through 3 give the data used to calculate radial metallicity gradients and other abundance patterns in this paper. In Table~1 we give the PN G number, heliocentric and Galactocentric distances (in kpc), the angular radius (in arcsec), the main morphology, and the PN progenitor type (OPPNe or YPPNe), if available, of the PNe in our study. 

In Table~2 we give the elemental abundances of He, C, N, O, Ne, Ar, selected by us from the references above (S method). 

In Table~3 we give the average abundances (A method), and their ranges, for all PNe with more than one abundance measurement of at least one of the elements He, C, N, O, Ne, or Ar. All abundances are from the above references. PNe with none or just one abundance measurement of each element in the reference above are not listed in table~3.

Tables 1 through 3 are published in their entirety only electronically.

\section{Radial $\alpha$-element gradients for the general  Galactic PN population}

In Figure~3 we show the elemental abundances of  oxygen, neon, and argon (from top to bottom) vs. Galactocentric distance ($R_{\rm G}$) for the PNe in our data set. In Table~4 we give the sample characteristics (Column 1), whether we have used selected (S) or average (A) abundances for the PNe (2), the sample size (3), the slope (4) and intercept (5) of the radial metallicity gradient derived by a simple fit. Unless noted otherwise, we use the SSV distance scale.  We find that the general Galactic $\alpha$-element gradient from PNe is shallow and negative, as already inferred in several earlier studies (see SH10 and references therein).  The slope for the general sample if $\sim$-0.02 dex kpc$^{-1}$ both for oxygen and neon, and significantly steeper ($\sim$-0.03 dex kpc$^{-1}$) for argon. Using the S or A method does not make a significant difference.

To test the stability of this general result, we recalculated the gradients with different assumptions. If we exclude bulge and halo PNe, defined as in SH10, we find gradient slopes of $\sim$-0.02  dex kpc$^{-1}$ both for O/H and Ne/H. We also found that the radial oxygen gradient for the general PN sample, such in Fig.~3, but using a different distance scale, varies by 10$\%$ at most, where the slope is flatter if using the brightness temperature distance scale \citep{VSZ1995}.

\section{Radial metallicity gradients as a function of age}

\subsection{Linear Radial oxygen gradients}\label{sec:lineargradient}

In Figure~4 we show the radial oxygen gradients derived for YPPNe (top) and OPPNe (bottom panel). In this plot we used S abundances, excluding those with known uncertainties larger than 0.5 dex.  Gradient slopes, obtained from simple linear fit,  are -0.015 and -0.027 for the OPPNe and YPPNe populations respectively (see Table~4). Linear fits can also be obtained using the {\it fitexy} routine, which take into account uncertainties in both axes \citep{NumRec}. A fit that includes the same data of Fig.~4 with uncertainties either measured, or assumed to be 0.2 dex, for all abundances, gives a $\chi^2$ fit probability of $q>0.9$ for both OPPNe and YPPNe, $\Delta{\rm log(O/H)}/\Delta R_{\rm G}$ (OPPNe)=-0.0013$\pm$0.005  dex kpc$^{-1}$ and $\Delta{\rm log(O/H)}/\Delta R_{\rm G}$ (YPPNe)=-0.023$\pm$0.008  dex kpc$^{-1}$, consistently with basic linear fits. The calculated uncertainties on the slopes, 0.005 and 0.008 dex kpc$^{-1}$ respectively for OPPNe and YPPNe, are smaller than the slope difference between the two population. We thus conclude that the OPPN radial oxygen gradient is significantly flatter than the YPPN one, indicating steepening of the gradient slope since Galaxy formation. 

If we (attempt to) exclude bulge and halo PNe from the sample, by the same approach used in SH10, we obtain a gradient slope for the old population of $\Delta{\rm log(O/H)}/\Delta R_{\rm G}$(OPPNe)=-0.019, and for the young population, $\Delta{\rm log(O/H)}/\Delta R_{\rm G}$(YPPNe)=-0.026. In general, by eliminating bulge and halo PNe from a sample, its metallicity gradient does not vary in a significant way.

By using OPPNe and YPPNe and measuring the radial oxygen gradients with Galactocentric distances based on a different distance scale, for example the brightness temperature distance scale \citep{VSZ1995}, we obtain $\Delta{\rm log(O/H)}/\Delta R_{\rm G}$ (OPPNe)=-0.0103 and $\Delta{\rm log(O/H)}/\Delta R_{\rm G}$(YPPNe)=-0.024. Using another distance calibration such as the surface brightness- to- physical radius correlation (e.g., \cite{frew16} does not change the gradients significantly either. 

All gradient plots showed so far are characterized by high scatter. More information can be derived by binning these plots in the Galactocentric distance direction, as shown in Figure~5. Here, the data of Figure~4 have been binned in Galactocentric distance bins of 1, 3, and 9 kpc, respectively, from top to bottom panels. Slopes of the YPPNe are always steeper than slopes of the OPPNe with the same binning, and intercepts are always larger for the young populations. The gradient slopes of Fig.~5 are $\Delta {\rm log(O/H)}/\Delta R_{\rm G}$ (YPPNe)=-0.034, -0.035, and -0.023 while $\Delta{\rm log(O/H)}/\Delta R_{\rm G}$ (OPPNe)=-0.015, -0.012, and -0.012 for the 1, 3, and 9 kpc binning respectively. In the figure it is clear that the gradients of the old and young populations are similar in the inner Galactic disk, and they diverge for large Galactocentric distances.

The comparison with other published gradients must take into account the radial extend and age range of the probes. For example, oxygen gradients have been measured by the APOGEE team using Galactic open clusters  \citep{cunha2016}. Most of the clusters have ages within 1-2 Gyr, thus they compare well with our YPPNe. \cite{cunha2016} found an oxygen gradient of -0.032$\pm$0.007 dex kpc$^{-1}$, which agrees with YPPNe gradient slope, within the uncertainties.

By applying a standard correction from O/H to Fe/H abundance gradients, iron gradient from YPPNe would be about 
-0.06 dex/kpc (2$\times$-0.03 dex/kpc), to compare with gradients measured on open clusters and field
stars around $\sim$-0.1 dex/kpc.
The reason why the PN gradient is shallower than other gradients
might be the distance range taken into account. The PN sample extends 
from the Galactic center to R$>$ 20 kpc, while open cluster are essentially observed at R$>$ 6 kpc. 
When the PN gradient is measured on a more restricted interval, (see $\S$6.2) the gradient 
is much steeper, and compatible with other tracers. 
The PN data is however too sparse to be able to measure the variation of the local gradient on the sample of
YPPNe, making it difficult to know if the gradient has flattened or steepened over the distance range 6-10 kpc, 
as seems to be the case for open clusters. 

From the gradients studied in this section we found a mild steepening of the gradient with time since Galaxy formation. The differences are small, but persisting in each experiment, across assumptions on distance scale, bulge and halo PN inclusion, and independent on the abundance choices.  

We explore the distance from the Galactic plane distribution of YPPNe and OPPNe, starting from the sample of Figure~4. While the scatter of the two populations is large, we find that the median distance from the Galactic plane of the OPPNe is twice that of the YPPNe, with $<|z|_{\rm YPPNe}>=265\pm269$ pc, while $<|z|_{\rm OPPNe}>=484\pm322$ pc (median values and standard deviations). Apparently,  YPPNe do not populate the thick disk, in agreement with stellar studies.

\subsection{Radial PNe step function}\label{sec:stepfunction}

The PN radial oxygen distributions shown so far may be best fitted with a step function rather than a continuous gradient.  In the linear fits to the radial distributions based on old progenitors  (Fig.~4, OPPNe), 
the fitted gradients tend to be systematically below the observed points in the 
range 5$<R_{\rm G}<$11 kpc.
In Fig.~4, most of the points with $R_{\rm G}<10$ kpc have oxygen above log(O/H)+12=8.5.  The
(few) points at $R_{\rm G}>$12-14 kpc are below the log(O/H)+12=8.5 line, meaning that there is a kind of a step function  
in the data. 
The distribution can be fitted by a flat gradient at $R_{\rm G}<$10 kpc and log(O/H)+12$\sim$8.6
(the slope is flat, with the fit giving a slope of 0.009 dex kpc$^{-1}$ in this region), and another flat gradient at $R_{\rm G}>$13.5, also with flat slope (-0.006 dex kpc$^{-1}$)  at log(O/H)+12$\sim$8.37, which, formally, represents a decrease of 0.065 dex kpc$^{-1}$ between 10 and 13.5 kpc.
In Figure~6 we show the OPPNe panel of Fig.~4, where we superimpose the gradients found by fitting the three regions, $R_{\rm G}<10$, 10$<R_{\rm G}<13.5$,  and  $R_{\rm G}>$13.5 kpc respectively.
  
Preliminary results on the OPPNe oxygen gradient slope with Gaia parallax distances (Stanghellini et al. 2018, in preparation), available only for R$_{\rm G}<10$ kpc, agrees with our result.

The step-like function seen on PNe is also observed in other tracers (although the distance scales for different probes may have relative biases). This is true in particular for open clusters, for which the gradient seem to be steep between about 7 and 10 kpc, 
and then flattens \citep[e.g.][]{lepine2011,frinchaboy2013,netopil2016, reddy2016}.
The line in between the two flat gradients of Figure~6 for OPPNe has a slope of roughly -0.033 dex kpc$^{-1}$
which, allowing for the standard factor of 2 for the conversion between O and Fe (see SH10 and references therein) would mean 
a gradient around -0.07 dex/kpc. 
Similar steep gradient have been found on field stars between 7 and 10 kpc \citep[e.g.][-0.09 dex kpc$^{-1}$]{bovy2014},
while for instance results from the APOGEE survey from \cite{hayden2015} found a gradient of about -0.05  dex kpc$^{-1}$ beyond R$\sim$6 kpc, with no 
indication of a flattening beyond 10~kpc. 
Fig.~7 of  \cite{hayden2015} shows that the peak of the thin disk MDF varies from +0.25 dex to -0.1 dex, 
or $\sim$ 0.35 dex decrease in metallicity between 6 and 10 kpc, a value similar to \cite{bovy2014}. 
Although with fewer points, the distribution of YPPNe and OPPNe suggests the same behavior, 
with the mean oxygen abundance below 10 kpc being significantly higher than the distribution beyond 
this limit, as is found on other data. 

Finally, it is interesting to note that OPPNe draws an homogeneous population 
with no significant gradient within 10 kpc, a characteristic which is also found on the thick 
disk population, which extends to the same limits, and has no detected gradient \citep[see][]{cheng2012}.

It is worth recalling that the Galactic PN distances may be slightly overestimated, since the SSV scale gives higher distances when compared to parallactic distances \citep{smith2015}. Using a different statistical scale will not change the results considerably.

\section{Vertical variations of the radial metallicity gradients}

\cite{cheng2012} noticed a lack of gradients in the thick disk. Their experiment with SEGUE stars show that by going to larger distances from the Galactic plane the [Fe/H] radial gradient decrease, until is almost flat in the highest altitude bin. Theoretical formulation of chemical evolution of the Galaxy by \cite{curir2014} has shown similar evolution (see their Fig.~4)

We studied our PN population as a function of the distance from the Galactic plane, and calculate oxygen gradients for both OPPNe and YPPNe, and also for the general sample, at different altitudes. We perform this experiment for PNe in the $6<R_{\rm G}< 16$ kpc space, to be consistent with \cite{cheng2012}'s analysis. 

In order to get useful statistics in each bin, we need larger $|z|$ bins than those used by \cite{cheng2012}. If we use all OPPNe with an oxygen abundance detection, the radial gradient for  $0.15<|z|<0.5$ kpc has slope $\sim$-0.04 dex kpc$^{-1}$, while the gradient calculated for $0.5<|z|<1.5$ kpc is $\sim$-0.02 dex kpc$^{-1}$. We thus recover flatter gradient at higher $|z|$, as for the SEGUE data, and obtained results that agree broadly with \cite{cheng2012}'s.

If we use the same conversion factor from O/H to Fe/H gradients as above, we obtain $\Delta(\rm {Fe/H)}/\Delta R_{\rm G}$=-0.08 dex kpc$^{-1}$ at $|z|\sim$0.35, and  $\Delta(\rm {Fe/H)}/\Delta R_{\rm G}$=-0.04 dex kpc$^{-1}$ at $|z|\sim 1.0$, which agree well the data of Figure~8 in Cheng et al. (2012).
Apart from precise gradient calculation, and considering the high scatter of our PN sample, there is a clear signature in the data that the radial metallicity gradient is flatter at high altitudes, consistent with the fact that the OPPN sample shows a flatter gradient.

A good summary of the various measurements available is given in \cite{xiang2015} (Fig. 15 therein),  which shows gradients
as a function of distance to the Galactic plane from various studies, compared with the results obtained with the LAMOST
spectroscopic survey by the authors.  

\cite{xiang2015} find that gradients of all ages but their oldest bin decrease with distance from the Galactic plane, from 
about [-0.10, -0.15] at z$<$0.5 kpc  to [-0.04, 0.0] dex/kpc at z$>$1.5 kpc. With a value of -0.08 dex/kpc at z$\sim$ 350 pc, 
our values are compatible with these values given the difference in R$_{\rm G}$ spanned by the various samples 
(the oldest stars in Xiang et al. range from R$\sim$7.5 to 10.5 kpc, while our PN sample
cover a distance interval approximately 5 times larger, from 0 to 15~kpc).
Similarly, at z$\sim$1~kpc, our value (-0.04 dex/kpc) is well within the range of the gradients obtained by Xiang et al. 2018, 
for stars younger than 11 Gyr, which is not surprising given the fact that our OPPN sample is probably contaminated by 
young objects no belonging to the thick disk, as already mentioned. 

The trend obtained from our OPPN stars is the same as in \cite{xiang2015} or \cite{hayden2015}: while the oldest (thick disk?) stars have a flat
gradient at large z, both the young and old stars have steeper gradients near the Galactic plane. The reason for this effect 
is not clear, but it cannot be excluded at this stage that samples of old stars in the plane are still significantly contaminated by 
younger objects, biasing the gradients.

\section{PN morphology and gradients}

By studying morphology of PNe in the different populations we found that the distribution of PN morphology are different for YPPNe and OPPNe. Of the 139 OPPNe of known shape,  $\sim60\%$ are either round or elliptical, while only 40$\%$ are bipolar, bipolar core, or point-symmetric \citep{iaccat} . On the other hand, we find that only 23$\%$ of YPPNe with known morphology are either round or elliptical, and the majority are either bipolar, bipolar core, or point-symmetric. While we do not have morphological classification for all PNe used in this study, these findings agree with the fact that young, high-mass AGB stars tend to produce asymmetric PNe \citep{stanghellini1993}.  Radial oxygen gradients for the symmetric (i.e., round or elliptical) PNe are marginally shallower than those for asymmetric (bipolar core, bipolar, and point-symmetric) PNe, with slopes respectively of -0.018 and -0.023 dex kpc$^{-1}$ for symmetric and asymmetric PNe respectively.

\section{Discussion}
A particularity of PNe as tracers is the very large distance range on which they can be sampled. As already mentioned, the radial 
baseline of our sample is 2-5 times more extended than for other tracers. Although our main discussion below concerns the thick disk, one may ask (and this is valid also for other galaxies) if we are justified to compare gradients of different epochs over such large distance range. If our sample of OPPN is dominated by thick disk objects, considering the thick disk on distance of about 10~kpc is justified because it has been found that the chemistry of the thick disk has been very homogeneous, see references in previous sections

The question becomes more complex concerning the thin disk, because there is a clearly a dichotomy of its chemical properties in the  inner and outer disks, and while the parenthood between the thick disk and the inner thin disk is conspicuous (Bovy et al. 2012, Haywood et al. 2013), the origin of the outer disk is not well understood. It is not clear, in particular, what role the thick disk had in shaping the gradient observed in the thin disk in various studies beyond R$_{\rm G}\sim$~6 kpc. In the discussion below, we leave this complex question apart, and concentrate more on evoking the reason of this dichotomy and on the consequences of a flat radial gradient in the thick disk. 

\subsection{The case of the Milky Way}

Unlike in SH10, the more complete data presented here allowed us to find
evidence of a step function in the radial distribution of oxygen abundance of old progenitor PNe vs. $R_{\rm G}$, see $\S$. \ref{sec:stepfunction}.
\cite{halle2015} \citep[see also][]{halle2018} proposed that the step feature observed on chemical abundances (metallicities of field 
stars or OC, oxygen abundances of PNe) 
is the effect of the presence of the outer Lindblad resonance of the bar (OLR) at about the same radius (7-10 kpc
from the Galactic center, see Dehnen, 2000, P{\'e}rez-Villegas et al., 2017). 
N-body simulations show that the bar allows stars in the disk to gain or loose angular 
momentum between resonances (hence to migrate by 'churning'), but are generally not allowed to cross the OLR of the bar, when the bar is 
the main asymmetry. The migration by churning essentially stops at the OLR and the stars cannot migrate further out by gaining angular momentum.
In the Milky Way disk, the OLR is located near the solar orbit (7-10 kpc from the Galactic center)
according to \cite{dehnen2000}, setting 
a barrier that separates the inner ($R_{\rm G}<$7 kpc) and outer ($R_{\rm G}>10$ kpc) disk, 
and that may be responsible for the dichotomy in the chemical properties of the inner and outer disk which is visible 
on local data \citep{haywood2013} and most clearly on in situ, APOGEE data, see \cite{hayden2015}.

Observations show that bars in MW-type galaxies are forming at epochs around $z=1-1.5$ \citep{sheth2008, melvin2014}.
If this is also the case for the MW, it means that this dynamical separation between the inner and outer 
disk could have been in place at the end of the thick disk formation, 
already 9 Gyr ago \citep[see][and Haywood et al. submitted]{haywood2016}. 
A long-lived bar and its OLR would then guarantee that these two regions would
remain essentially separated in the last 8-9 Gyr, explaining the observed  dichotomy and the step function
also reflected in the distribution of PNe  in chemical abundances. 

Radial metallicity gradients are key constraints for chemical evolution models. Gradients can be modified by various physical processes 
whose contribution we must try to assess.
Within 10 kpc of the Galactic center, the data shows that the oxygen gradient is essentially flat for OPPNe. While it is not clear 
how much the OPPN sample represents the thick disk, this is reminiscent of the flat gradient that has been found 
for this population with the SEGUE data (Cheng et al. 2012).
It has been argued that even if there had been a gradient in the thick disk, mixing of stars (by either 
blurring or churning), would have erased it. 
It must be reminded that this is not the case, for several reasons discussed in \cite{haywood2015}. 
A metallicity gradient at a given epoch means that, at the corresponding age, in the thick disk, 
stars of different metallicities and $\alpha$-elements are formed. 
Yet, it is observed that at any given age in the thick disk, metallicities and $\alpha$-elements have a very small dispersion 
\citep{haywood2015,haywood2016}, implying an absence of gradient, and hence an absence of inside-out process in this population.
We comment further on this result in the next section.

\subsection{Clues to disk formation}

In nearby, external spirals, the evolution of metallicity gradients has been inferred by comparing those from the populations of H~II regions and PNe. Even if they belong to different stellar evolutionary paths, these emission-line probes are observed and analyzed identically, often within the same multi-object observed frame, thus the contrast between the resulting gradients is very credible.  PNe and H~II regions have been observed directly to derive radial oxygen gradients in a small sample of galaxies. All galaxies where PNe and H~II regions have been both studied with direct abundances, M33: \cite{magrini2009, magrini2010}; M81: \cite{stanghellini2010,stanghellini2014}; and NGC300: \cite{stasinska2013}, showed moderate (or null) steepening of the radial oxygen gradient with time. 
Are gradients indeed steepening, or are they flattening, with time evolution? The local galaxies all seem to indicate that most likely gradients are not highly variable in the times probed by observations.

In Figure~7 we compare the radial oxygen gradients that we determined in this paper for old and young progenitor PNe with \cite{gibson2013}'s models. We plot the radial oxygen gradient slopes, 
$\Delta$log(O/H)/$\Delta$R$_{\rm G}$, vs. redshift $z$, from \cite{gibson2013}'s models as black lines in the figure (broken lines are models with enhanced feedback). Models by Moll{\'a} (2014) would be very similar to those by \cite{gibson2013} without enhanced feedback. Other simulations have confirmed the effect of feedback as the driving force of turbulence in disk at high redshifts, see in particular
\cite{ma2017}, who find a very shallow gradient. 

In the figure we also plot several Galactic and extragalactic data sets, not necessarily homogeneous, for discussion purposes rather than direct comparison with the models. For redshifted galaxies, we plot the slopes of radial oxygen gradients (or range thereof) inferred from H~II regions, vs. $z$. In these cases, gradients are calculated from abundances of oxygen in H~II regions obtained through strong line analysis,  and not from direct abundances. 

For local galaxies, ordinates (gradient slopes) are determined from direct abundances, while abscissae are calculated from the age of the probe used in the gradient determination. Gradients from H~II regions in local galaxies have z=0, and those from PNe are plotted at a z representing the age of formation of the PN progenitors. For PNe in the  present study, we have assumed that OPPNe probe stars older than $t_{\star}>$7.5 Gyr, and YPPNe probe stars younger than $t_{\star}<$1 Gyr. Redshifts are derived from population ages (lookback times) from a simple cosmological model, where $\Omega_{\rm matter}$=0.3, $\Omega_{\lambda}$=0.7, h=0.7. We plot the gradient slopes that we have found for these two populations, marking both these points with open circles in Fig.~7. The arrows indicate that these are respectively a lower and upper limit to the population age for OPPNe and YPPNe respectively.

The vertical lines at z$\sim$3.3 and  1.2 mark respectively the range of radial oxygen gradients of the galaxy samples analyzed  by 
\cite{cresci2010}, and in \cite{sanchez2014}.  The crosses and triangle are from analysis of lensed galaxies, thus probably not homogeneous with our data sets, due to the difficulty of reconstructing the actual galactic gradient. The dots are direct abundances of PNe and H~II regions in M33 and M81.  

We find a qualitative agreement between data and models especially in the enhanced feedback group, which at this point seem to give a good indication about which models may better represent the majority of stellar populations in star forming galaxies.

\subsection{The thick disk evolution}

Although in Fig.~7 the circle at $t_{\star}=7.5$  (corresponding to log(1+z)$\sim$0.35) represents OPPNe, with progenitors predominantly in the thick disk, it  certainly cannot represent only the thick disk and must be contaminated by younger population.
The fact that OPPNe at $R_{\rm G}>$14 kpc have a mean oxygen abundance lower than those at R$<$10 kpc (hence producing a gradient)
probably means that they belong to an outer disk, which is thought to be a different population than 
the 'chemically defined' thick disk, which is limited essentially to $R_{\rm G}<$10~kpc \citep{cheng2012}. 
In this case, restricting the measured gradient to $R_{\rm G}<$10 kpc, where the thick disk is 
more dominant, makes more sense, and we can consider the gradient to be essentially flat (see Section \ref{sec:stepfunction}). 
A flat metallicity gradient is a natural outcome of simulations with feedback, which induce sufficient turbulence 
to mix metals on large scales in less than a Galactic time. Hence, in \cite{gibson2013}, the MaGICC simulations
(strong feedback) have a gradient limited to about -0.02  dex kpc$^{-1}$ at the epoch of thick disk formation (redshift $\sim$2).
Similarly, in \cite{ma2017}, 
the disk between 1 and 2 kpc from the Galactic plane has a gradient that varies from -0.01 to +0.01  dex kpc$^{-1}$, passing 
by no gradient at $|z|$=1.5 kpc.
It is interesting to note how this point to the same picture advocated in \cite{haywood2013, haywood2015} and 
Haywood et al. (submitted), where we argued that the chemical evolution of the thick disk must have been 
very homogeneous, and incompatible with the standard assumed inside-out evolution. Inside-out formation 
scenario means that star formation operates at a faster rate towards the inner disk, inducing radial gradients 
and dispersion in metallicity and $\alpha$ abundances  at a given age  \citep[e.g.][]{minchev2014, kubryk2015} that are not observed \citep[][Haywood et al., submitted]{haywood2015}.

\section{Conclusions}

By selecting a large sample of Galactic PNe, and using the best available abundances, we found a flat general radial oxygen gradient, with slope $\sim$-0.02 dex kpc$^{-1}$. We use a novel age determination method for the progenitors of the PNe, based on the comparison of chemical yields and AGB evolutionary models. 

We found that the radial oxygen gradient slope is almost twice as steep for PNe whose progenitors are younger than 1 Gyr (YPPNe), than for those whose progenitors are older than 7.5 Gyr (OPPNe), indicating  a mild steepening of the oxygen gradient with time evolution.

By applying the standard conversion between oxygen and iron gradients for YPPNe we can compare our results with those from open clusters and field stars we found that the PN gradients are shallower, probably due to the different Galactocentric distance ranges of the PNe vs. compared stellar samples. If we limit PN gradients to more restricted interval, the gradients become much steeper, and compare well with the general stellar population.

The radial oxygen gradients for PNe with old progenitors can be well modeled by a step function, with a flat radial gradient inside 10 kpc and outside 13.5 kpc from the Galactic center. This step-like function of the radial distribution of metals in the Galaxy has been seen with other tracers. By studying the OPPN sample as a whole, it is clear that they are mostly thick disk probes (although the sample is contaminated), showing a much flatter radial metallicity gradient at higher altitudes, which is consistent with the SEGUE data results by Cheng et al. (2012). By restricting the OPPN sample to R$<$10 kpc, where the thick disk is more dominant, we observe an essentially flat gradient, which agrees with \cite{gibson2013}'s models with strong feedback at the time of thick disk formation.
\clearpage

\begin{deluxetable}{lrrrll}

\tablecaption{Input Parameters of PNe}

\tablehead{ 
\colhead{PN~G}&  \colhead{Dist}& \colhead{R$_{\rm G}$}& \colhead{$\theta$}& \colhead{Morph.}& \colhead{Prog.} \\
&[kpc]& [kpc]& [arcsec]& &\\
}

\startdata
    000.1+17.2 &  14.51 &   5.86 &   0.90 &    E &   ... \\
    000.1-01.1 &   8.63 &   0.63 &   1.90 &    E &   ... \\
    000.1-02.3 &   5.58 &   2.42 &   4.50 &  ... &   ... \\
    000.1-05.6 &   7.92 &   0.12 &   6.20 &    E &   ... \\
    000.2-01.9 &   8.80 &   0.80 &   2.50 &    B &   ... \\
    000.3+12.2 &   3.97 &   4.12 &   4.20 &  ... &  OPPN \\
            \enddata
   \end{deluxetable}

\begin{deluxetable}{lrrrrrr}

\tablecaption{Abundances, selected (S)}

\tablehead{ 
\colhead{PN~G}&  \colhead{He/H\tablenotemark{a}}& \colhead{C/H}& \colhead{N/H}& \colhead{O/H}& \colhead{Ne/H}&\colhead{Ar/H}\\
}

\startdata
 000.3+12.2 &  10.99 $\pm$0.01& $\dots$& 7.68 $\pm$0.25& 8.50 $\pm$0.07& 7.82 $\pm$0.18 & 5.61 $\pm$0.10 \\
    000.7+03.2 &  11.20 $\pm$0.04& $\dots$& 8.46 $\pm$0.21& 8.63 $\pm$0.27& 7.73 $\pm$0.19 & 6.63 $\pm$0.20 \\
    001.4+05.3 &  11.07 $\pm$0.08& $\dots$& 7.35 $\pm$0.25& 8.42 $\pm$0.99& $\dots$& 6.35 $\pm$0.62 \\
    001.6-01.3 &  10.98 $\pm$0.03& $\dots$& 8.45 $\pm$0.11& 8.27 $\pm$0.05& $\dots$& 6.29 $\pm$0.06 \\
    001.7-04.6 &  11.09 $\pm$0.03& $\dots$& 7.76 $\pm$0.55& 8.76 $\pm$0.16& 8.11 $\pm$0.06 & 6.53 $\pm$0.09 \\
           \enddata
       \tablenotetext{a}{All abundances are in the format log(X/H)+12, and the uncertainties are logarithmic}
   \end{deluxetable}

\begin{deluxetable}{lrrrrrr}

\tablecaption{Abundances, averages (A)}

\tablehead{ 
\colhead{PN~G}&  \colhead{He/H\tablenotemark{a}}& \colhead{C/H}& \colhead{N/H}& \colhead{O/H}& \colhead{Ne/H}&\colhead{Ar/H}\\
}

\startdata
    002.0-13.4 &  11.04$\pm$0.01& $\dots$& 7.97$\pm$0.36& 8.68$\pm$0.02& 8.03$\pm$0.03 & 6.32 \\
    002.2-02.7 &  11.00$\pm$0.04& $\dots$& 7.77$\pm$0.40& 8.35$\pm$0.05& 7.67$\pm$0.04 & 5.82$\pm$0.28 \\
    002.2-09.4 &  11.08$\pm$0.04& $\dots$& 8.43$\pm$0.01& 8.92$\pm$0.14& 8.38 & 6.85$\pm$0.15 \\
    002.7-04.8 &  11.19$\pm$0.04& $\dots$& 7.91$\pm$0.54& 8.84$\pm$0.35& 8.32 & 6.72$\pm$0.40 \\
    003.1+02.9 &  11.11$\pm$0.01& $\dots$& 8.52$\pm$0.21& 8.69$\pm$0.02& 8.06$\pm$0.07 & 6.43$\pm$0.20 \\

         \enddata
       \tablenotetext{a}{All abundances are in the format log(X/H)+12, and the uncertainties are logarithmic}
   \end{deluxetable}

\begin{deluxetable}{llrrr}

\tablecaption{Galactic metallicity gradients from PNe}

\tablehead{ 
\colhead{Sample}&  \colhead{abun.}& \colhead{N$_{\rm PNe}$}& \colhead{Slope}& \colhead{Intercept} \\
&&&[dex kpc$^{-1}$]& [log(X/H)+12]\\
}

\startdata
All PNe \\
O/H& A& 300& -0.016& 8.685\\
O/H& S& 295& -0.018& 8.689\\
O/H\tablenotemark{a}& S&		248&			-0.021&		8.715\\

Ne/H& A& 251& -0.019& 8.116\\
Ne/H& S& 243& -0.018& 8.100\\
Ne/H\tablenotemark{a}& S& 208& -0.020& 8.134\\

Ar/H& A& 207& -0.029& 6.426\\
Ar/H& S& 223& -0.029& 6.426\\
Ar/H\tablenotemark{a}& S& 179& -0.030& 6.438\\
\hline
YPPNe\\
O/H&	A&		57&			-0.025&		8.730\\
O/H& S&   54&  -0.027& 8.727\\
\hline
OPPNe\\
O/H&	A&		189&			-0.016&		8.702\\
O/H&	S&		171&			-0.015&		8.702\\

O/H\tablenotemark{b}&  S&  46&      -0.048&  9.117\\
O/H\tablenotemark{c}&  S&    36 &      -0.021&    8.709\\
        \enddata
       
\tablenotetext{a}{Excluding bulge and halo PNe.}
\tablenotetext{b}{Gradient calculated only for PN in the $6<R_{\rm G}<16$, and $0.15<|z|<0.5$ kpc domain.}
\tablenotetext{c}{Gradient calculated only for PN in the $6<R_{\rm G}<16$, and $0.5<|z|<1.5$ kpc domain.}

\end{deluxetable}

\clearpage

\acknowledgements 
We acknowledge the thorough review of an earlier version of this paper, and many helpful suggestions, by an anonymous Referee. 
L.S. is grateful to Drs. Katia Cunha, Laura Magrini, Mark Dickinson, Beatrice Bucciarelli, Mario Lattanzi, and Paolo Ventura for scientific discussions, and to Drs. Michel Dennefeld and 
Valerie de Lapparent for science conversations and for their hospitality at the IAP in the Spring of 2016, when most of this work has been developed.

\begin {figure}
\plotone {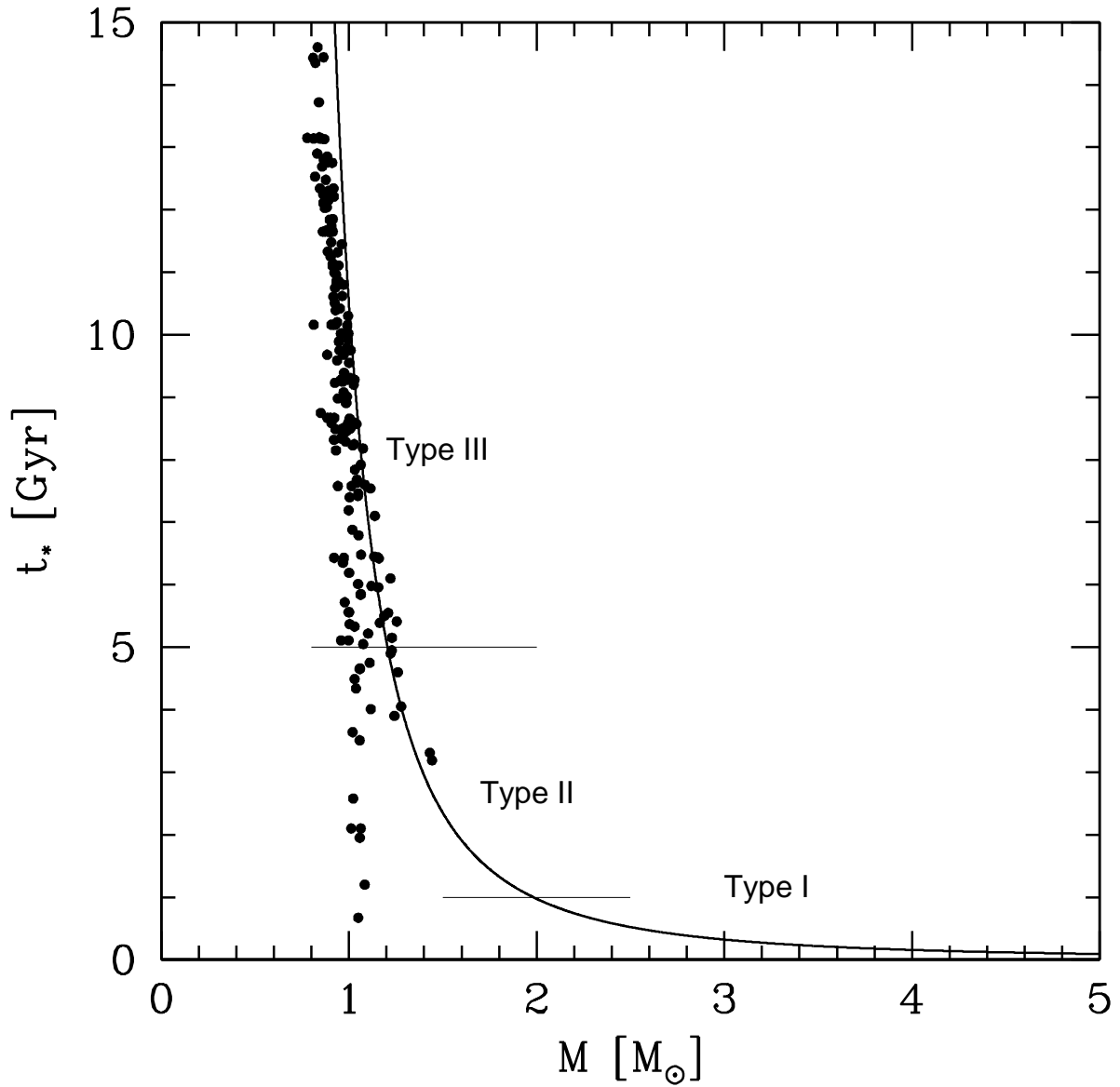}
\figcaption{The heavy solid line is the stellar age (in Gyr) vs. mass (in M$_{\odot}$) relation from the FCT \citep{maraston1998}. The line has been divided into three sectors (by light horizontal lines) determining the realms of Type~I, Type~II, and Type~III PN progenitors (see SH10 for definitions). The solid circles are stellar ages and masses of our selection  of stars (see text) from \cite{ramirez2013}.}
\end {figure}

\begin {figure}
\plotone {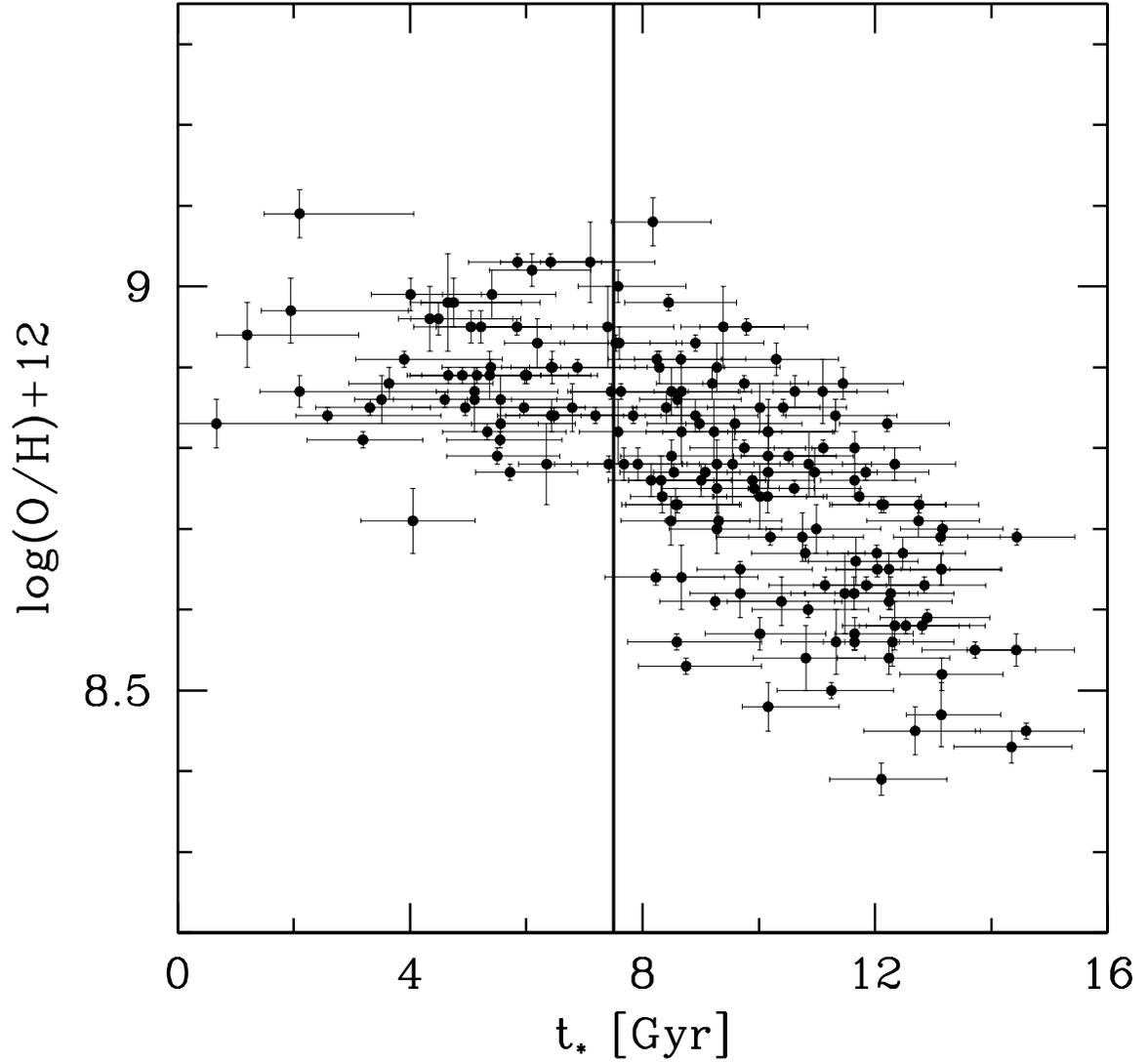}
\figcaption{The oxygen abundances vs. stellar age, as derived from  \cite{ramirez2013} stellar data and our selection (see text). Uncertainties are in the original reference. The solid vertical line marks $t_{\star}=7.5$ Gyr.}
\end {figure}

\begin {figure}
\plotone {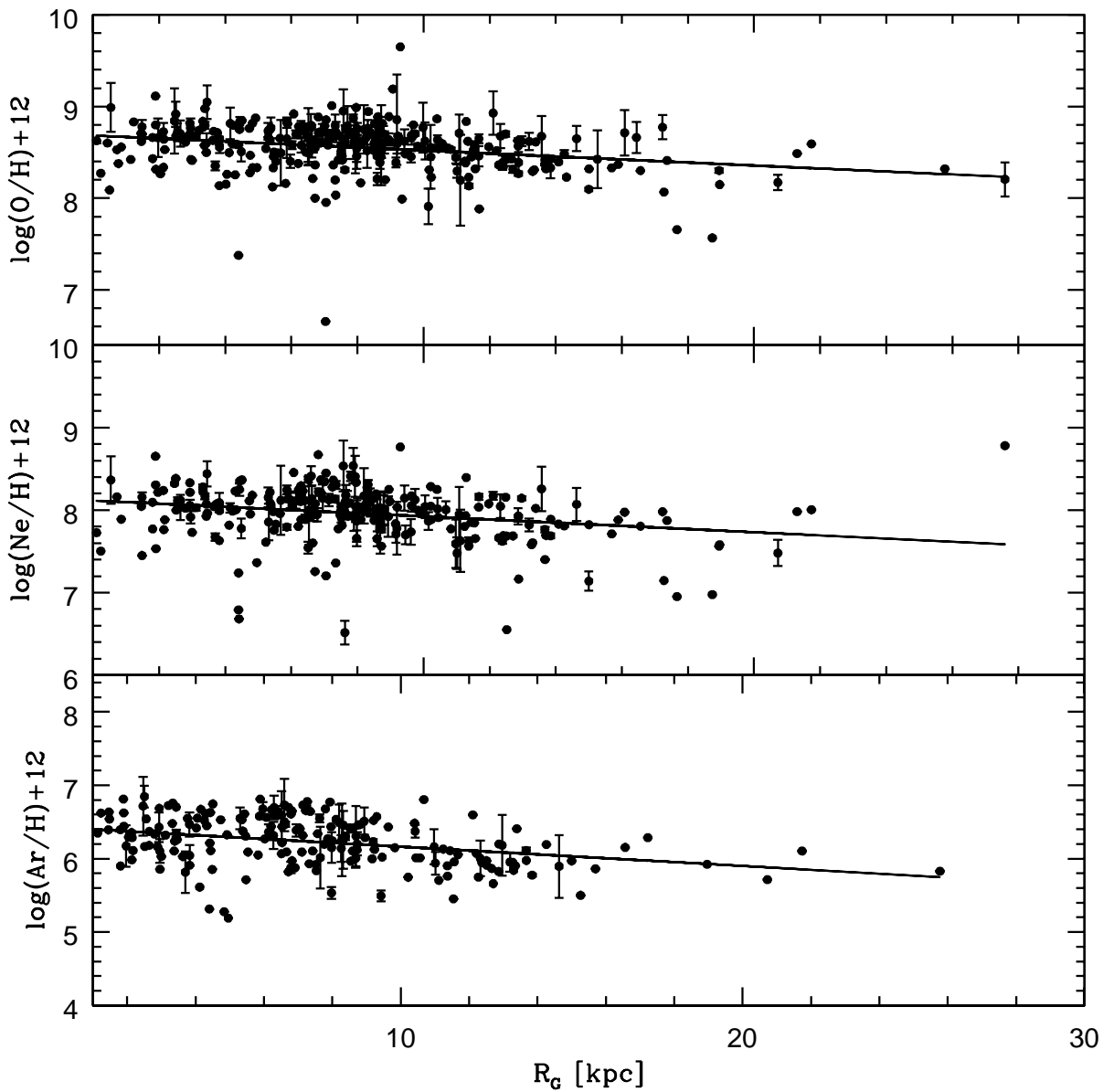}
\figcaption{Radial metallicity gradients based on Galactic PN abundances. In these panels we show the oxygen (upper panel), neon (middle panel) and argon (lower panel) gradients for Galactic PNe selected as described in the text. For each PN we used the average elemental abundance for each element, or A, as described in the text. The gradient slopes are given in Table~4.}
\end {figure}

\begin {figure}
\plotone {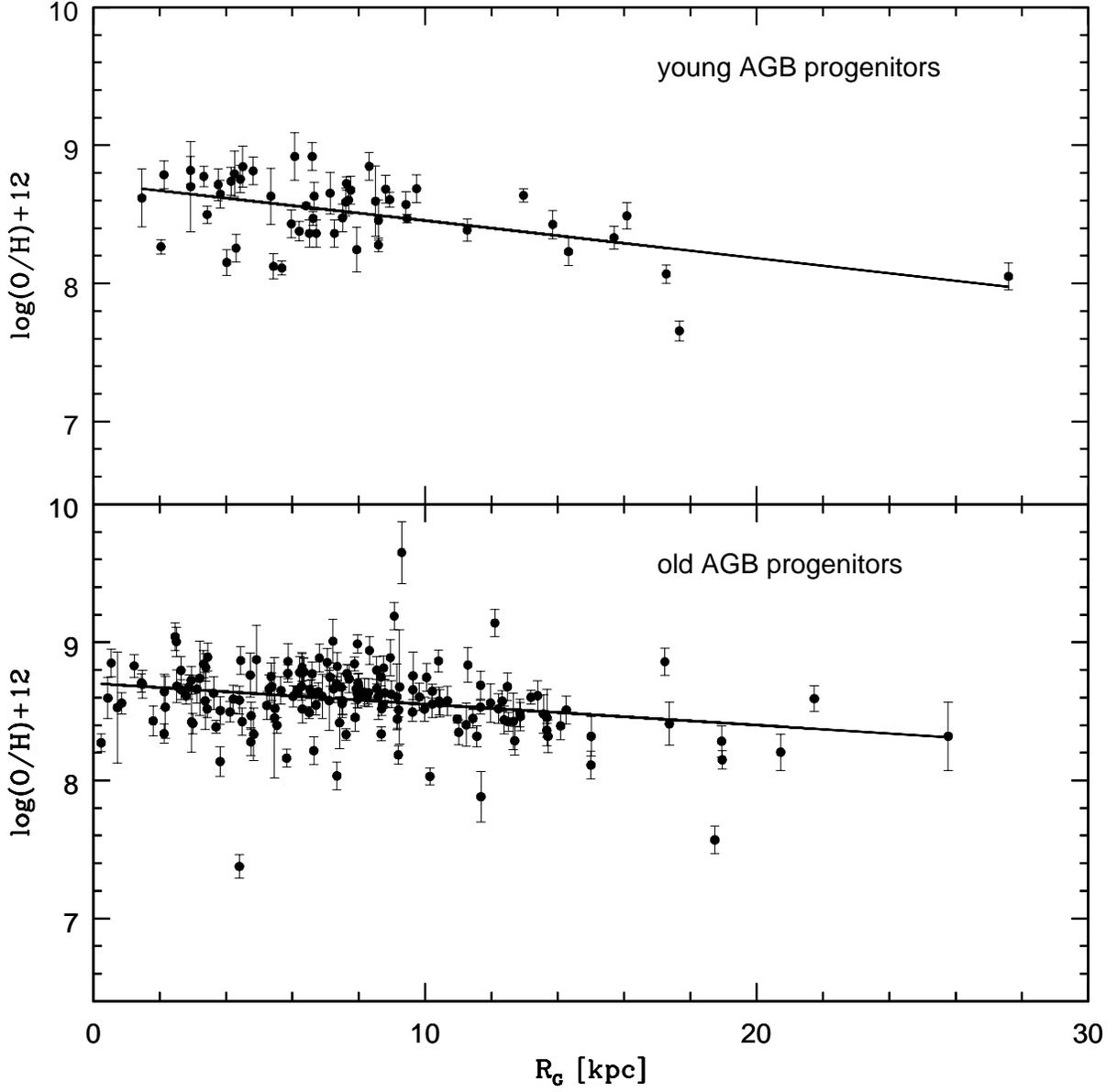}
\figcaption{Radial oxygen gradients for PNe with old (upper plot, OPPNe) and young (lower plot, YPPNe ) progenitors. The abundances are from the S method. }
\end {figure}

\clearpage 

\begin {figure}
\plotone {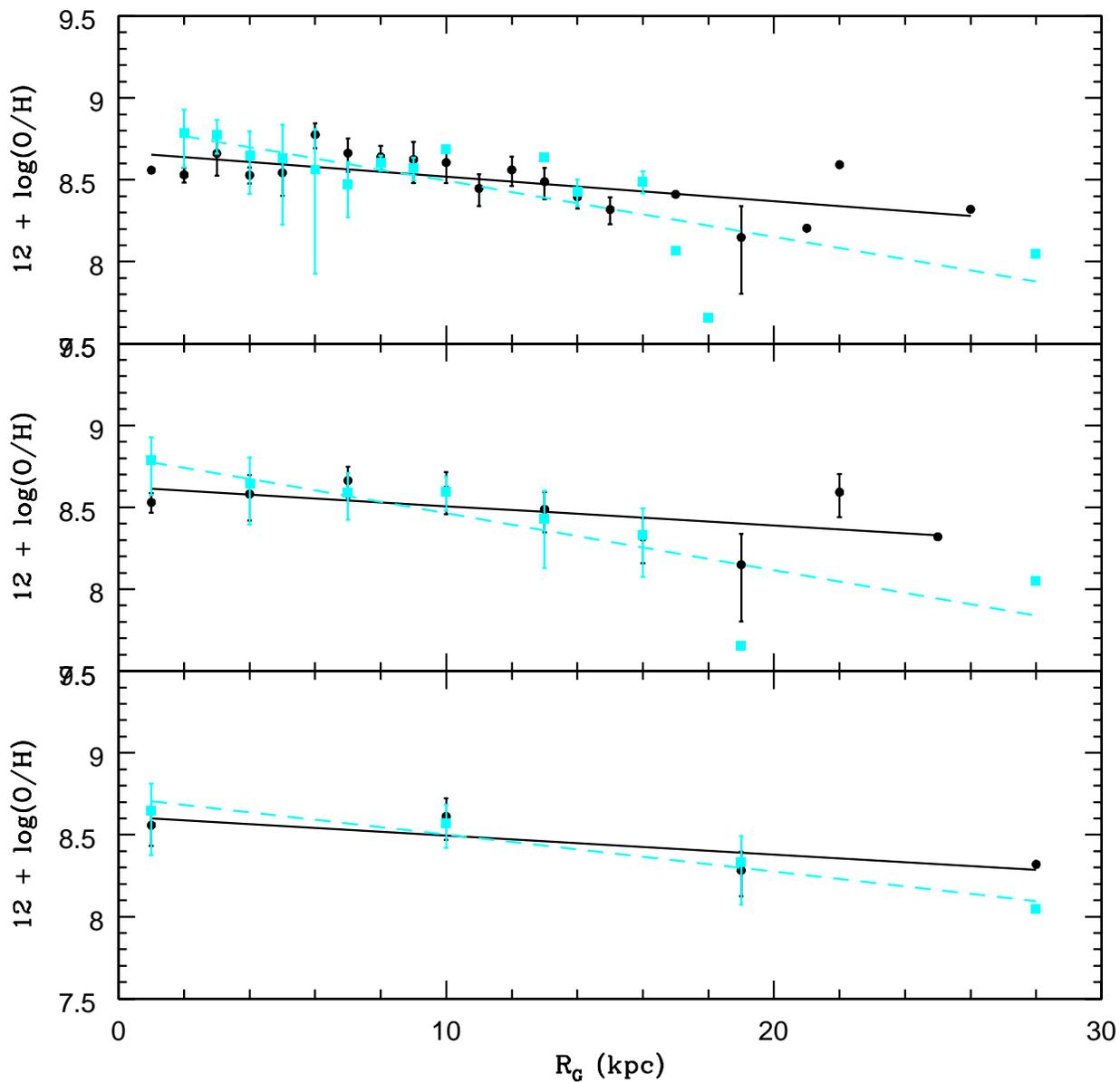}
\figcaption{Binned gradients for OPPNe (filled circles, and solid lines) and YPPNe (filled squares, and broken lines). Oxygen abundances have been averaged over bins of 1 (top panel), 3 (middle panel), and 9 (bottom panel) kpc. Error bars represent abundance ranges within each bin. }
\end {figure}

\begin {figure}
\plotone {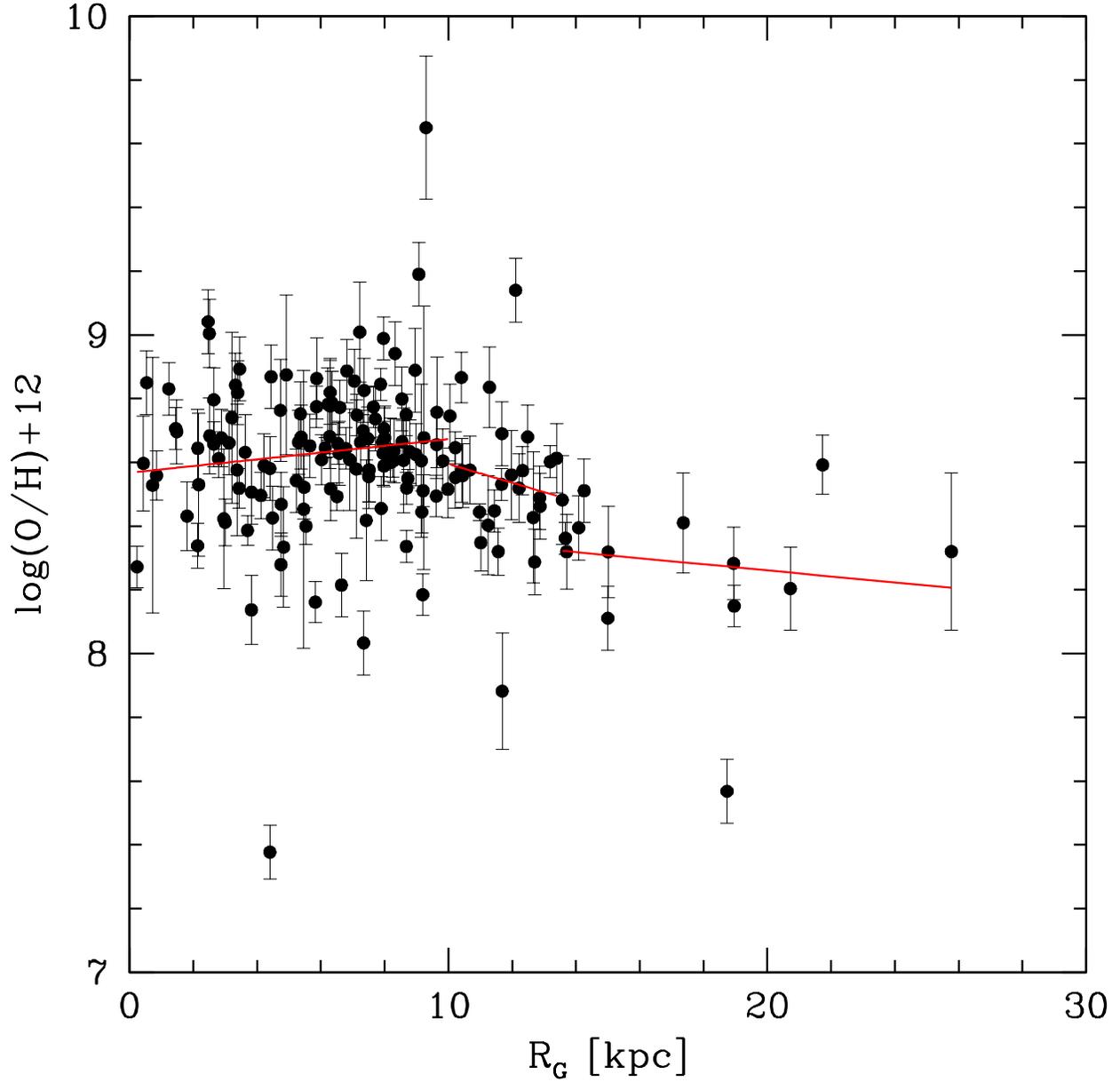}
\figcaption{Radial distribution of oxygen abundances in OPPNe as in Fig.~4, lower panel. Here we overplot the inner and outer flat gradients, from the fit of the data with R$_{\rm G}<10$, $>$13.5 kpc, and in between, see text. }
\end {figure}
\clearpage

\begin {figure}
\plotone {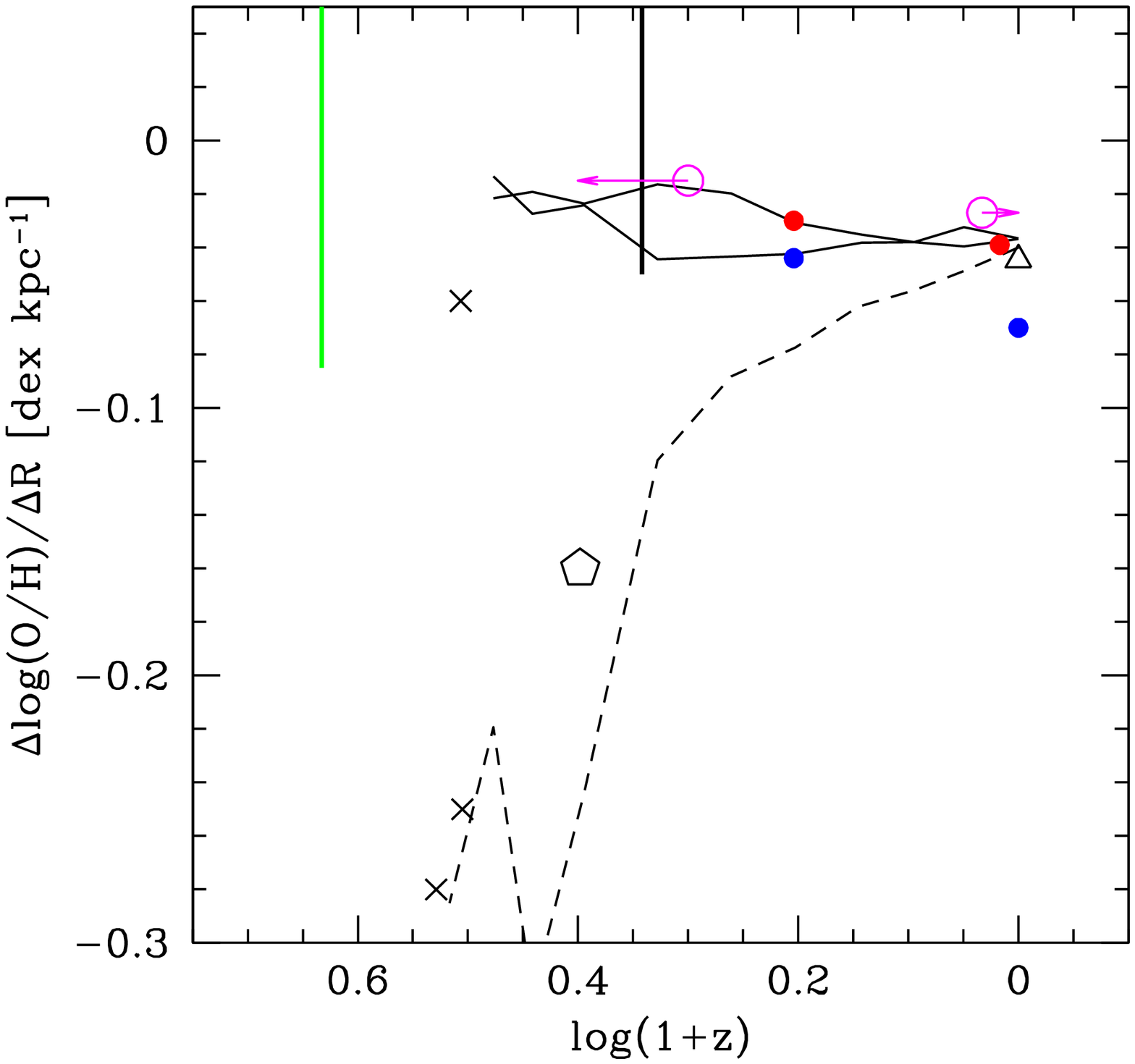}
\figcaption{The radial oxygen gradient from star forming galaxies vs. redshift. Lines are representative of inside-out chemical galaxy evolution models in a cosmological context, solid lines: with enhanced feedback, broken lines: no feedback \citep{gibson2013}.  Triangle: gradient evolution from Galactic H~II regions \citep{balser2011}. Red dots: M33 PN and H~II region gradients \citep{magrini2009, magrini2010}; Blue dots: Gradients from PNe and H~II regions in M81 \citep{stanghellini2010, stanghellini2014}; pentagon: \cite{yuan2011}; crosses: \cite{jones2013};  vertical lines: ranges of gradient slopes from \cite{cresci2010} and \citep{sanchez2014}. Open circles: OPPNe and YPPNe, this work.}
\end {figure}

\clearpage

\end{document}